\documentclass[twocolumn,pra]{revtex4}
\usepackage{amstext}
\usepackage{amsmath}           
\usepackage{amssymb}           
\usepackage{graphicx}          
\usepackage{latexsym}
\newcommand{\ket}[1]{\vert#1\rangle}

\begin{document}
\title{Berry's phase in Cavity QED: proposal for observing an effect of field quantization}
\author{A. Carollo$^{\diamond\ddag}$, M. Fran\c{c}a Santos$^{\diamond}$, and  V. Vedral$^{\diamond}$}
\address{$^{\diamond}$Optics Section, The Blackett Laboratory,
Imperial College, London SW7 2BZ, United Kingdom \\
$^\ddag$INFM, unit\`{a} di ricerca di Palermo, via Archirafi 36,
90123 Italy}

\begin{abstract}
Geometric phases are well known in classical 
electromagnetism and quantum mechanics since the early works of
Pantcharatnam~\cite{pac} and Berry~\cite{berry}. Their origin relies on the
geometric nature of state spaces and has been studied in many different systems
such as spins, polarized light and atomic physics~\cite{wil}. Recent works have
explored their application in interferometry and quantum computation~\cite{jvec,fal}.
Earlier works suggest how to observe these phases in single quantum systems adiabatically
driven by external classical devices or sources, where, by classical, we mean any
system whose state does not change considerably during the
interaction time: an intense magnetic field 
interacting with a spin $1/2$, or a birefringent medium interacting
with polarized light~\cite{tom}-\cite{tom6}. Here we propose
a feasible experiment to investigate quantum effects in these phases, arising
when this classical source drives not a single quantum system,
but two interacting ones. In particular, we show how to observe a signature of
field quantization through a vacuum effect in Berry's phase. To do so, we describe the
interaction of an atom and a quantized cavity mode altogether driven by an external
quasi-classical field.
\end{abstract}
\maketitle

The matter-light interaction as well as the quantum nature of light itself have 
been subjects of many experimental investigations in the last
two decades, specially in the Cavity QED domain~\cite{haroche1,haroche12,walther,walther2}. Furthermore,
it has recently been shown that the quantization of the electromagnetic field can affect
the geometric phase acquired by an evolving two-level system, if this system is made to interact 
with the quantized field~\cite{angelo}. In this letter, we propose an experiment that allows one
to observe such an effect in Cavity QED.

Our system uses two dispersive Jaynes-Cummings 
interactions~\cite{jcm}, in the $\Lambda$ Raman configuration, between the atomic 
levels, two orthogonally polarized cavity modes (+ for the right and - for the left polarization)
and an auxiliary external field $\vec{E}(t)=\varepsilon_0[(\cos\frac{\theta}{2}
e^{i(\frac{\phi}{2})}\hat{\epsilon}_+ + \sin\frac{\theta}{2}
e^{-i(\frac{\phi}{2})}\hat{\epsilon}_-)e^{i\omega t} + \it{c.c.}]$ inside a
microwave cavity (see fig1). The quantum system to be observed
consists of the electronic levels of the atom dressed by the
cavity modes. The external field, injected transversely in the
cavity, works as our external driving source.

Due to the dispersive one-photon interaction, the two upper
electronic levels of the atom act only as virtual states allowing
for the two-photon coupling between levels $|1\rangle$ and $|2\rangle$. The
polarization of the auxiliary field determines how each cavity
mode couples to the effective transition $|1\rangle \rightarrow |2\rangle$.
For large detuning, i.e. $\delta \gg {g, \Omega}$, the whole system evolves
according to the effective Hamiltonian~\cite{marcelo}
\begin{multline}
H_{eff}= \hbar\frac{\Omega^2}{\delta} \hat{\sigma}_{22} +
\hbar\frac{g^2}{\delta}(\hat{a}^{\dag}_+\hat{a}_+ +
\hat{a}^{\dag}_-\hat{a}_-) \hat{\sigma}_{11} \\+
\lambda [(\cos\frac{\theta}{2}e^{i\frac{\phi}{2}}\hat{a}_+ +
\sin\frac{\theta}{2}e^{-i\frac{\phi}{2}}\hat{a}_-) \hat{\sigma}_{21} + h.c.],
\end{multline}
where $\hat{a}_+$ ($\hat{a}_-$) denotes the annihilation operator
of the right (left) circularly polarized quantized mode and we assumed states
$|1\rangle$ and $|2\rangle$ to have the same parity. We assume, for 
the sake of simplicity, that the atom couples with the same strength to each 
polarization, both with the classical and the quantized modes $g_+=g_-=g=\Omega$.
We also consider $\Omega_+=\cos\frac{\theta}{2}e^{i\frac{\phi}{2}}\Omega$ and
$\Omega_-=\sin\frac{\theta}{2}e^{-i\frac{\phi}{2}}\Omega$. Finally, we define 
$\lambda = \frac{g\Omega}{\delta}$ as the effective coupling constant of the 
whole system.

The adiabatic rotation of the polarization of the injected (external) driving
field produces two simultaneous effects; the usual one, which
is related to the geometric phase $\gamma/2$ acquired by this quasi-classical
(intense) field, where $\gamma$ is the solid angle 
associated to the surface described by its
polarization vector in the Poincar\'e sphere (fig2). The other effect,
much more subtle, is the slow variation of the 
effective Hamiltonian parameters $\theta$ and
$\phi$, and, therefore, of the internal structure of the
atom-quantized modes interaction.

In particular, if this rotation is such that the final Hamiltonian
is the same as the original one, except for the phase $\gamma/2$,
then a quantum Berry phase can be observed in the joint state of
the atom and the quantized modes. According to~\cite{angelo},
for a resonant Jaynes-Cummings interaction, state $|1\rangle
|n-1\rangle |m\rangle$ ($|2\rangle |n\rangle
|m\rangle$) acquires a geometric phase
$e^{-i\gamma /2(n-m+1/2)}$ ($e^{i\gamma/2(n-m+1/2)}$), where $\ket{n}$ and 
$\ket{m}$ are the Fock states of modes $\hat{a}_+$ and $\hat{a}_-$, respectively.

Note that even if both cavity modes are initially in the vacuum
state, there is still a phase shift $e^{i\gamma/4}$ in the atomic
state $|2\rangle$, while state $|1\rangle$ remains unchanged.

This geometric phase shift can be observed through the usual Ramsey
interferometric method (fig3)~\cite{haroche1}. The atom is prepared in the first
Ramsey zone and put into a symmetric superposition of levels 
$|1\rangle$ and $|2\rangle$, while mode $a_+$ is prepared in
some arbitrary state and mode $a_-$ is prepared in the vacuum. The
initial polarization of field $\vec{E}(t))$ is chosen so that only mode
$a_+$ interacts with the atom. The external field is turned on and
the whole system is allowed to interact for a time $\tau \gg
1/\lambda$, during which the polarization of the external field is
rotated at a much slower rate compared to $\lambda$ until it is
back to its original value. After that, the atom undergoes a $\pi/2$
rotation in the second Ramsey zone and is detected by ionization.

If the initial state of mode $a_+$ is also the vacuum, then
$H_{eff}$ is resonant (for $\Omega = g$) and the probability to
detect the atom in level $|2\rangle$, after an integral number of
Rabi flips inside the cavity, and the $\pi/2$ pulse of the second
Ramsey zone, is given by $P_2=(1-\cos\gamma/4)/2$. This
corresponds to a $\gamma/4$ geometric phase shift, which is a
surprising effect, comparing to the usual $\gamma/2$ classical result.
The atom-cavity mode system splits into doublets
except for a lonely ground state, which serves as an internal
reference frame for the measurement of the global geometric phases
acquired by the remaining dressed states. This is only possible
due to the quantum behaviour of the cavity mode.

This becomes clear, if the experiment is repeated but, now, a
coherent state $|\alpha\rangle$ is initially injected in mode
$a_+$. Then, the probability to measure the atom in the excited
state for the same $\gamma$, is given by ~\cite{kike}
\[2P_2=(1-e^{-|\alpha|^2})(1-\cos\gamma/2)
+e^{-|\alpha|^2}(1-\cos\gamma/4).\]
From this equation
it is clear that only for $|\alpha\rangle = 0$, we have the
surprising $\gamma/4$ ratio. As the intensity of the initial
coherent state increases, the observed phase shift goes to
the well expected $\gamma/2$ semi-classical limit. In this calculation, we
assume a slightly different setup in which the dynamical effects 
can be eliminated~\cite{kike}.

In Fig. 2a, we show the phase shift in the probability to find the
atom in level $|2\rangle$ $P_2$ for a induced geometric phase
$\gamma = \pi$ due to the rotation of the polarization of the
driving field. In Fig. 2b we show $P_2$ as a function of different
initial coherent states in the cavity field, for a fixed
interaction time $\tau = 10/\lambda$, and the same phase shift
$\gamma = \pi$.

Typical values for the one photon vacuum Rabi frequency are
$g/2\pi \simeq 50 kHz$~\cite{haroche1,haroche2}. Choosing the same coupling
for the classically driven transition $\Omega/2\pi \simeq 50 kHz$,
and a detuning $\delta = 3\Omega$, $\lambda/2\pi \simeq 15 kHz$,
which means that the atom-field system can perform approximately
$10$ complete Rabi cycles during an effective interaction time of
$0.6 ms$. Times of interaction of this order are achievable in
microwave cavities for atomic velocities of the order of $10 m/s$
and are within typical decaying times ($1$ ms) for the cavities used
in~\cite{haroche1}. According to our calculations, in these
experimental conditions, the non-ideal adiabatic rotation of the
polarization produces errors of the order of $5\%$ in $P_2$,
which are smaller than the ones produced by other sources, like
detectors inefficiency, cavity field decay and so on.

Our work suggests that the geometric phase still has additional surprising 
features. It also shows the versatility and power of the simple 
combination of the quantized cavity mode manipulated by an 
externally injected quasi-classical field. This opens a number of further 
possibilities for research, such as for example the effect 
of the phase on the collapses and revivals of atomic population in the
Jaynes-Cummings model~\cite{collapse}. 

\noindent {\bf Acknowledgements} This research was supported by EPSRC, Hewlett-Packard, 
Elsag spa and the EU. MFS acknowledges the support of CNPq.

Correspondence should be addressed to v.vedral@ic.ac.uk.


\vspace*{0.3cm}

\newpage

Figure 1. The experimental proposal. The Rydberg atoms cross cavity C one at a time;
the cavity mode is initially prepared in some weak coherent state and a circularly 
polarized intense field $\vec{E}(t)$ drives the atom-cavity mode system.
The atomic levels are shown in fig 1b. For large detunings $\delta \gg g,\Omega$,
the two lower levels interact effectively through a two-photon process; for simplicity,
we consider $g=\Omega$, where $\Omega_+=\cos\frac{\theta}{2}e^{i\frac{\phi}{2}}\Omega$ and
$\Omega_-=\sin\frac{\theta}{2}e^{-i\frac{\phi}{2}}\Omega$; in R1, the atoms are prepared 
in symmetric quantum superpositions of levels 1 and 2, and subsequently perform an integral 
number of Rabi flips inside C. During this interaction in C, the polarization 
of the external field is slowly rotated so as to complete a closed loop in the Poincar\'{e} 
sphere (as in fig. 2); the atoms are finally rotated back in R2, and detected in I.

Figure 2. A closed loop described by the polarization vector of the external 
driving field. The (classical) geometric phase generated in this way is equal 
to one half of the solid angle enclosed by the path (equal to $\pi$ in this figure). 
Note that the same picture can be used to represent the evolution 
of a quantum two-level system whose state space has the same structure 
as a two dimensional sphere (known as the Bloch sphere). In our paper we 
combine these two pictures with the quantization of the field to produce 
and measure a new phase effect. The evolution of the joint atom and 
the quantized field state can no longer be represented in this simple way.

Figure 3. Experimental prediction. (a) probability to measure the atom in state 
$|2 \rangle$, for a fixed number of Rabi flips inside the cavity and different
relative phases $\xi$ between R1 and R2; the green curve is the caliber curve for the
Ramsey interferometry, and the red curve represents the situation in which a 
geometric phase shift of $\pi/4$ is induced in the system, according to the
proposed experiment. This phase shift corresponds to the solid angle $\gamma = \pi$
described in fig 2. Part (b) represents the effective phase shift for different coherent 
states prepared in the cavity. We see that for the vacuum states (i.e. a 
coherent state of amplitude $0$) the shift would be equal to $\pi/4$, 
and it would increase with the coherent state amplitude converging to 
the expected (semi-classical) value of $\pi/2$. Note that the convergence 
is very fast, and for amplitudes of $2$ and higher the semi classical result 
is in a very good agreement with the fully quantized calculation.

\end{document}